\documentclass[amsmath,amssymb,twocolumn,notitlepage,nofootinbib,superscriptaddress,floatfix,eprint,aps,prl]{revtex4-2}

\usepackage[utf8]{inputenc}
\usepackage[english]{babel}
\usepackage{graphicx}
\usepackage{float}
\usepackage{color}
\usepackage{lineno}
\usepackage{tabularx}
\usepackage[usenames,dvipsnames,table]{xcolor}
\usepackage[normalem]{ulem}
\usepackage{amsthm}
\usepackage{eucal}
\usepackage{times,bbm,amsmath,amssymb}
\usepackage{epsfig,color}
\usepackage{xcolor}
\usepackage{bm}
\usepackage{booktabs}
\usepackage{float,siunitx}
\usepackage[caption = false]{subfig}
\usepackage{booktabs}

\usepackage{sidecap}
\usepackage[scaled=.8]{couriers}
\usepackage{pstricks}
\usepackage{multirow}
\usepackage{placeins}
\usepackage{relsize}
\usepackage{pst-grad,bm}
\usepackage{epigraph}
\usepackage{gensymb}
\usepackage{longtable}
\usepackage{soul}
\usepackage{ulem}
\usepackage{xr-hyper}
\usepackage{hyperref}
\usepackage{cleveref}
\hypersetup{
    colorlinks = true,
    allcolors=blue
}
\usepackage{layouts}
\usepackage{acronym}
\usepackage{physics}
\usepackage{easyReview}
\usepackage{soul}
\usepackage{amsmath,amssymb,amsfonts}
\usepackage{textcomp}
\usepackage{physics}
\usepackage[ruled,linesnumbered]{algorithm2e}
\usepackage{verbatim}
\usepackage{tabularx}
\usepackage{booktabs} 
\usepackage{microtype}
\microtypecontext{spacing=nonfrench}
\microtypesetup{
protrusion={true,nocompatibility},
activate={true,nocompatibility},
tracking=true,
kerning=true,
spacing={true}
}
\usepackage[all=normal,floats=tight,mathspacing=tight,wordspacing=tight,paragraphs=normal,tracking=tight,charwidths=tight,mathdisplays=tight,sections=normal,margins=normal]{savetrees}
\usepackage[style=american,autopunct=true]{csquotes}
\usepackage{placeins} 

\definecolor{blue}{rgb}{0,0,1}
\definecolor{grey}{rgb}{0.6,0.6,0.6}
\definecolor{myurlcolor}{rgb}{0,0,0.7}
\definecolor{myrefcolor}{rgb}{0.8,0,0}
\definecolor{purple}{RGB}{128,0,128}
\definecolor{ultramarine}{RGB}{63, 0, 255}
\definecolor{medblue}{RGB}{0, 0, 100}
\definecolor{googleblue}{RGB}{34, 0, 204}
\definecolor{panblue}{RGB}{0,24,150}
\definecolor{carmine}{RGB}{150, 0, 24}
\definecolor{gray}{RGB}{150, 150, 150}

\begin{document}

\title{Experimental verifiable multi-client blind quantum computing on a Qline architecture}

\author{Beatrice Polacchi}
\affiliation{Dipartimento di Fisica - Sapienza Universit\`{a} di Roma, P.le Aldo Moro 5, I-00185 Roma, Italy}

\author{Dominik Leichtle}
\affiliation{School of Informatics, University of Edinburgh, 10 Crichton Street, EH8 9AB Edinburgh, United Kingdom}

\author{Gonzalo Carvacho}
\affiliation{Dipartimento di Fisica - Sapienza Universit\`{a} di Roma, P.le Aldo Moro 5, I-00185 Roma, Italy}

\author{Giorgio Milani}
\affiliation{Dipartimento di Fisica - Sapienza Universit\`{a} di Roma, P.le Aldo Moro 5, I-00185 Roma, Italy}

\author{Nicol\`o Spagnolo}
\affiliation{Dipartimento di Fisica - Sapienza Universit\`{a} di Roma, P.le Aldo Moro 5, I-00185 Roma, Italy}

\author{Marc Kaplan}
\affiliation{VeriQloud, 13 rue Victor Hugo, 92 120 Montrouge, France}

\author{Elham Kashefi}
\email{elham.kashefi@lip6.fr}
\affiliation{Laboratoire d’Informatique de Paris 6, CNRS, Sorbonne Université, 75005 Paris, France}
\affiliation{School of Informatics, University of Edinburgh, 10 Crichton Street, EH8 9AB Edinburgh, United Kingdom}

\author{Fabio Sciarrino}
\email{fabio.sciarrino@uniroma1.it}
\affiliation{Dipartimento di Fisica - Sapienza Universit\`{a} di Roma, P.le Aldo Moro 5, I-00185 Roma, Italy}

\begin{abstract}
The exploitation of certification tools by end users represents a fundamental aspect of the development of quantum technologies as the hardware scales up beyond the regime of classical simulatability. Certifying quantum networks becomes even more crucial when the privacy of their users is exposed to malicious quantum nodes or servers as in the case of multi-client distributed blind quantum computing, where several clients delegate a joint private computation to remote quantum servers, such as federated quantum machine learning. In such protocols, security must be provided not only by keeping data hidden but also by verifying that the server is correctly performing the requested computation while minimizing the hardware assumptions on the employed devices.
Notably, standard verification techniques fail in scenarios where the clients receive quantum states from untrusted sources such as, for example, in a recently demonstrated linear quantum network performing multi-client blind quantum computation.
However, recent theoretical results provide techniques to verify blind quantum computations even in the case of untrusted state preparation.
Equipped with such theoretical tools, in this work, we provide the first experimental implementation of a two-client verifiable blind quantum computing protocol in a distributed architecture. The obtained results represent novel perspectives for the verification of multi-tenant distributed quantum computation in large-scale networks.
\end{abstract}

\maketitle

\textbf{\textit{Introduction.}}
Quantum technologies promise to outperform classical devices in several tasks ranging from cryptography \cite{bennett2014quantum,ekert1991quantum,ekert1992quantum,pirandola2020advances} and computation \cite{deutsch1992rapid,shor1994algorithms,arute2019quantum,zhong2020quantum,aaronson2019complexity} to randomness generation \cite{acin2016certified,liu2018device,herrero2017quantum}.
However, analogously to the development of classical computers, near-term quantum devices are expected to be available to end users on the cloud \cite{cacciapuoti2019quantum,wehner2018quantum,kimble2008quantum}.
This opens several privacy issues on the users' side, who must be equipped with tools to keep their data private, \textit{blindness} of the computation, but also to check whether the quantum server they are employing is behaving honestly and is correctly performing the desired computations, \textit{verification} of the computation.
Certifying whether a quantum device is performing the correct operation is a challenging and widely investigated task \cite{agresti2021experimental,vsupic2020self,yang2014robust,gheorghiu2019verification,d2023machine,pironio2016focus,eisert2020quantum,acin2016certified}.
On one hand, the challenges mainly come from the required resource overhead, e.g. for quantum tomography procedures, or from the need for partial assumptions about the inner details of the quantum hardware under investigation. This is often seen as a major drawback, since, once quantum technologies are available to general users, the latter may completely ignore the technical details of the hardware they are using and may need certification protocols that rely on minimal assumptions about it, i.e. the so-called device-independent framework \cite{pironio2016focus}.
On the other hand, such certification procedures represent a key ingredient for realizing a secure quantum internet \cite{kimble2008quantum,cacciapuoti2019quantum,wehner2018quantum}.
Their importance becomes even more evident when users perform delegated tasks, such as delegated blind quantum computing (BQC) \cite{liu2023public,kapourniotis2022framework,ma2022qenclave,li2021blind,gheorghiu2017rigidity,aharonov2017interactive,gheorghiu2015robustness,perez2015iterated,hajduvsek2015device,hayashi2015verifiable,morimae2014verification,morimae2013blind,sueki2013ancilla,mantri2013optimal,reichardt2012classical,morimae2012blind,dunjko2012blind,fitzsimons2017unconditionally,leichtle2021verifying}, a class of protocols which allow users with minimal quantum resources to delegate hard computations to powerful remote quantum servers while keeping hidden input and outcome data, as well as the calculation itself. 
Due to the flourishing of federated protocols, such as federated machine learning \cite{konevcny2016federated,yang2019federated}, several multi-client versions of BQC have been theoretically proposed \cite{kapourniotis2023asymmetric,shan2021multi,qu2021secure,ciampi2020secure,kashefi2017multiparty} and recently experimentally demonstrated in a two-client setting \cite{polacchi2023multi}.
The theoretical proposals also included the possibility of verifying the computation, which was also demonstrated in different settings \cite{drmota2023verifiable,huang2017experimental,greganti2016demonstration,barz2013experimental,barz2012demonstration}, with the assumption that the clients owned trusted single-qubit sources.
The protocol presented in \cite{polacchi2023multi}, instead, is based on a linear quantum network structure, known as Qline \cite{doosti2023establishing}, featuring an untrusted quantum source that distributes qubits along linear quantum channels. Thanks to the versatility of this architecture, the clients only need single-qubit polarization rotation devices. This led to an advantage with respect to the original universal BQC protocol \cite{broadbent2009universal}, where clients were required to own single-qubit sources too.
However, the use of the Qline architecture triggered a fundamental question regarding the verifiability of the computation in this scenario, since the single-qubit sources are not owned by the clients and are, therefore, untrusted.
Notably, the issue of verifying BQC with untrusted sources of quantum states was recently addressed in \cite{kashefi2024verification}, where the authors demonstrate that, if the clients can also perform bit-flip operations together with $z-$rotations, they can verify that the server is correctly performing the computation, even without trusting the qubit source.

In this Letter, we present the first experimental implementation of a verifiable multi-client BQC protocol in a two-client setting, where the clients receive single qubits from an untrusted source. 
In detail, we exploit the versatile adaptive photonic platform introduced in \cite{polacchi2023multi}, where the clients are connected through fiber links and perform single-qubit transformations sequentially over a linear quantum channel. The main advantages of such a platform reside in the linear disposition of the clients along the quantum network and in the presence of a trusted third party (TTP) that secures the classical communication between the clients and the server, therefore reducing the time latency on the server's side.
We then devise this photonic platform to be capable of performing the verifiable blind quantum computation protocol introduced in \cite{kashefi2024verification}, by equipping the clients also with suitable polarization rotation devices and using some of the protocol rounds as test rounds to check for the server honesty.
Our results can be further extended to more parties at arbitrary distances, thus providing a step forward toward the realization of a secure and scalable distributed quantum cloud infrastructure.
We present the key ideas of the protocol with a two-client and two-server scenario.

\begin{figure}
    \centering
    \includegraphics[width=\columnwidth]{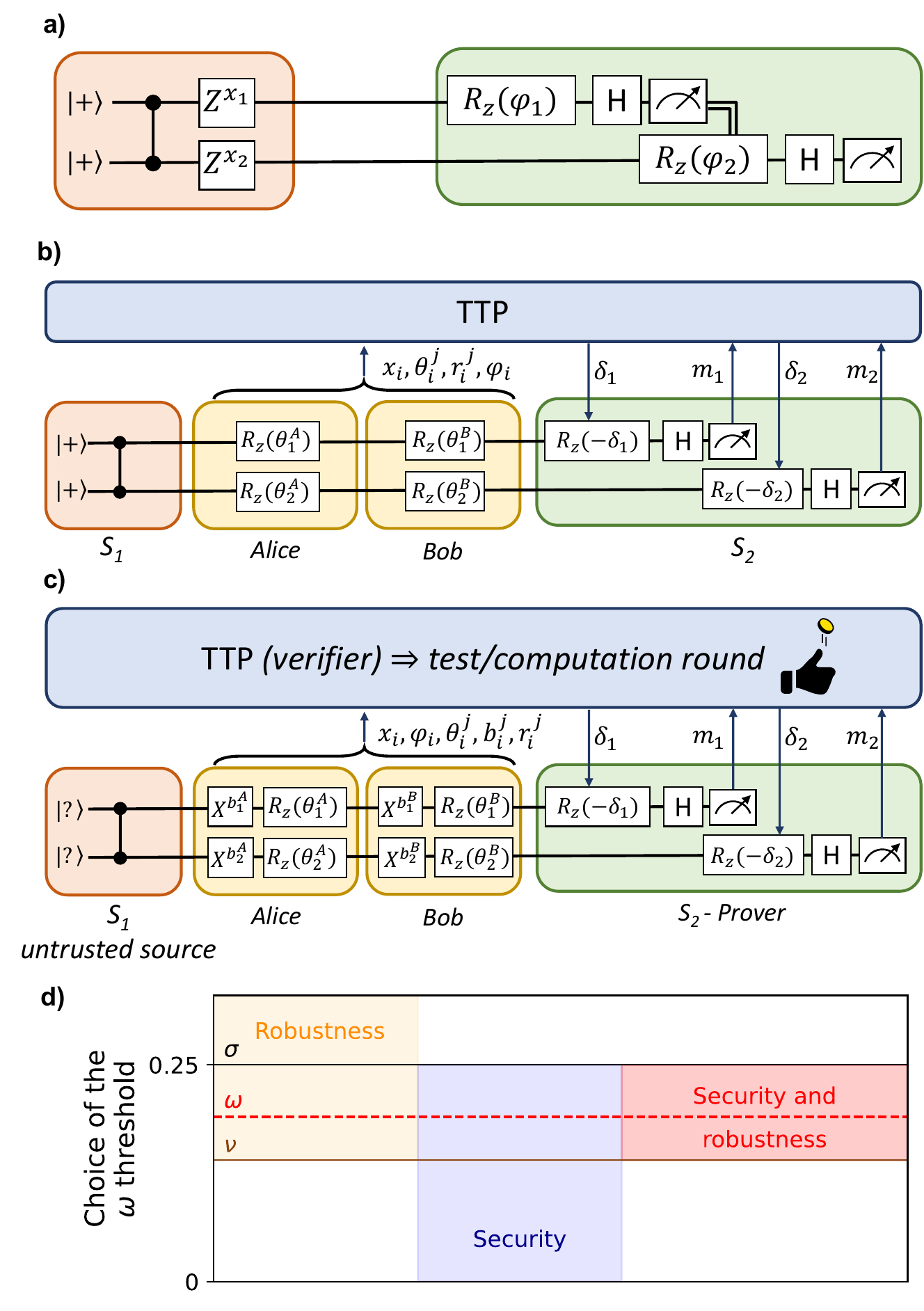}
    \caption{\textbf{Conceptual scheme of verifiable multi-client BQC with two clients and two Qlines. a)} Target circuit without any masking. \textbf{b)} Non-verifiable BQC with two clients. \textbf{c)} Modification of the scheme in \textbf{b)} to perform verifiable BQC with two clients.
    \textbf{d)} Choice of the threshold of tolerated failed test rounds $\omega$. This threshold must be chosen in the interval between the maximum theoretical threshold for security, $\sigma$, and the maximum expected level of hardware noise, $\nu$.}
    \label{fig:scheme}
\end{figure}

\textbf{\textit{Multi-client blind quantum computing.}}
We suppose that Alice and Bob own only single-qubit transformation devices, and want to delegate the joint computation shown in Fig.~\ref{fig:scheme}a to a remote quantum server while keeping their private data hidden.
In our protocol, we suppose that the clients desire to perform the classical function described by the two-qubit measurement pattern $\{ \phi_1, \phi_2 \} = \{ \frac{\pi}{2}, \frac{\pi}{2} \}$. 
The two clients respectively insert two classical input bits, $x_1$ and $x_2$.
To mask the computation data, Alice and Bob carry out the multi-client BQC protocol introduced in \cite{polacchi2023multi} and depicted in Fig.~\ref{fig:scheme}b.
However, some modifications must be applied to this protocol for the client to carry out the verification of the computation. 
Let us first briefly revise the non-verifiable BQC protocol introduced in \cite{polacchi2023multi}. 
Apart from the clients, the other parties involved in the protocol are (i) an untrusted source of qubits, that we will call server S$_1$, (ii) a trusted third party (TTP) that orchestrates and secures classical communications between the clients and the server, and (iii) a remote quantum server that performs the computation, indicated as S$_2$. 
In the ideal scenario where the source is behaving honestly, it distributes maximally entangled bipartite states along two Qlines, of the form:
\begin{equation}
    \ket{\psi} = \frac{1}{2} \left( \ket{00} + \ket{01} + \ket{10} - \ket{11} \right)
\label{eq:generated_state}
\end{equation}
The clients sequentially apply the following one-time pad transformation to the two qubits: $R_z(\theta_1^j) \otimes R_z(\theta_2^j) $, where the angles $\theta_i^j$ are randomly drawn in the set $\mathcal{A} = \{ 0,\pi/4,...,7\pi/4 \}$. We use $i=1,~2$ as the qubit index and $j=A,~B$ as the client index.
Moreover, each client chooses two random bits $r_i^j$ to hide the outcome of the computation.
The clients then communicate their secret parameters $\theta_i^j, r_i^j$ to the TTP. 
Let us define the quantities $\theta_i =  \theta_i^A + \theta_i^B$ and $r_i = r_i^A \oplus r_i^B$, for the $i$-th qubit.
The resulting quantum state is then sent to server S$_2$.
From now on, the clients and S$_2$ only communicate classically, through the TTP.
The protocol requires one round for each qubit, and, at the $i-$th round, the TTP computes the blind measurement angle $\delta_i$ according to the formula:
\begin{equation}
\delta_i = \theta_i  + x_i\pi + (-1)^{m_{i-1}^{\text{true}}}\phi_i + r_i \pi
\label{eq:delta_non_ver}
\end{equation}
where $m_0^{\text{true}} = 0$ and $m_1^{\text{true}} = m_1 \oplus r_1$.
Both outcomes are corrected by the TTP according to $m_i^{\text{true}} = m_i \oplus r_i$ at the end of the protocol. The protocol is repeated many times and the outcome of the computation is chosen through majority vote among all rounds.

We stress that, despite the blindness (privacy) of the clients' data being ensured even in the case of malicious source and server, the correctness of the computation outcome is never certified. Indeed the clients must assume that the qubit source is sending the desired quantum states and that the server is correctly computing the desired function. This represents a strong assumption, especially in the absence of any distributed fault-tolerant architecture. In the next section, we describe a modified version of this protocol in which the clients can drop these assumptions and verify the server behaviors.

\textbf{\textit{Verifiable multi-client blind quantum computing.} }
To allow for verification of the computation in the presence of an untrusted source of states, we need to apply two main modifications to the previously introduced protocol. 
A conceptual illustration of the modified protocol is depicted in Fig.~\ref{fig:scheme}c.
First, we need to repeat the experiment $n$ times, $m$ of which are used as test rounds, while $n-m$ as computational rounds \cite{kashefi2024verification,leichtle2021verifying}.
We notice that test rounds introduce only a polynomial overhead in terms of the round complexity of the protocol, thus preserving the efficiency and scalability of the original protocol.
The results from the test rounds are used to decide whether the computational results can be accepted as correct or not. 
The choice between test or computation is performed run-by-run from the TTP.

A second modification is due to the need of changing the state preparation procedure. In detail, the clients now need to randomly prepare the received qubits according to the transformation $R_z(\theta_1^j) X^{b_1^j} \otimes R_z(\theta_2^j) X^{b_2^j}$, as shown in Fig.~\ref{fig:scheme}c. The bits $b_i^j$ are randomly chosen by the clients to perform the verification protocol \cite{kashefi2024verification} and secretly communicated to the TTP.
In computation rounds, the blind measurement angle $\delta_i$ at the $i-$th round must be computed in an adaptive way by the TTP according to the formula given in Eq.~\eqref{eq:delta_non_ver}, where the angles $\theta_i$ are substituted with the angles $\theta'_1 = (-1)^{b_1^B}\theta_1^A + \theta_1^B + (b_{2}^A \oplus b_{2}^B) \pi$ while $\theta'_2 = (-1)^{b_2^B}\theta_2^A + \theta_2^B + (b_{1}^A \oplus b_{1}^B) \pi$.
In test rounds, instead, the target algorithm of the clients is the stabilizer measurement $Y \otimes Y$.
Indeed, since the outcome of such measurements is known, in this way the clients can detect eventual malicious deviation on the server side \cite{leichtle2021verifying,kashefi2024verification}.
In practice, this means that the second measurement is fixed to $\delta_2 = \theta_2'  + x_2\pi + \phi_2 + r_2 \pi$.

Test rounds can fail because of malicious deviations from the ideal behavior of the server or because of overly noisy hardware.
While the \textit{security} of the protocol ensures blindness and verifiability in the case of malicious deviations, even if the server behaves honestly, real-life hardware is affected by noise.
Hence, the \textit{robustness} property ensures that in the realistic, noisy case the protocol accepts with high probability.
As a core parameter of the protocol, the clients need to choose a threshold for the number of tolerated failed test rounds~\cite{leichtle2021verifying}.
The need for security and robustness yields two different bounds on the tolerable test failure rate $\omega$ in the protocol. On one hand, the security threshold $\sigma$, an upper bound on $\omega$, is given by the following formula (see Theorem~1 of \cite{leichtle2021verifying}):
\begin{equation}
    \sigma = \frac{1}{k}\cdot\frac{2p-1}{2p-2},
\end{equation}
where $k$ is the number of different types of test rounds and $p$ is the inherent error probability of the target BQP computation. In our configuration $\sigma = 0.25$.
Only for a choice of $\omega < \sigma$, the security of the protocol is guaranteed.
On the other hand, the expected magnitude of hardware noise implies a natural lower bound $\nu$ on $\omega$ to obtain robustness, where $\nu$ is the expected maximum test failure rate on the used hardware (see Theorem~2 of \cite{leichtle2021verifying}).
Such a parameter depends on the features of the untrusted qubit source, the clients' rotation devices and the server's hardware and can be, for example, a public property of a given service provider.
However, we stress the important fact that the protocol security does not depend on the choice of $\nu$. Indeed, it is only provided as an indication to the clients about the lower-bound on the choice of the $\omega$ threshold that must be taken into account in order to achieve sufficient robustness to the noise present in the setup, \textit{i.e.}, a reasonable rate of protocol round acceptance.
In other words, a careless choice of $\nu$ (and hence of $\omega$) could at most cause perpetuate aborts of the protocol, but could not lead to the clients accepting wrong results and therefore to a breach of security.

Once the clients have chosen the threshold of tolerated failed test rounds $\omega$, they perform the protocol and analyze the results as follows.
If the recorded error fraction during test rounds, $\epsilon$, is such that $\epsilon \leq \omega$, the clients perform a majority vote on the computation rounds.
If the recorded error fraction $\epsilon$ is such that $\epsilon > \omega$, the protocol aborts.
These parameters are tunable depending on the desired level of privacy and robustness from the clients' perspective, analytically providing the required number of protocol rounds.

\begin{figure}
    \centering
    \includegraphics[width=\columnwidth]{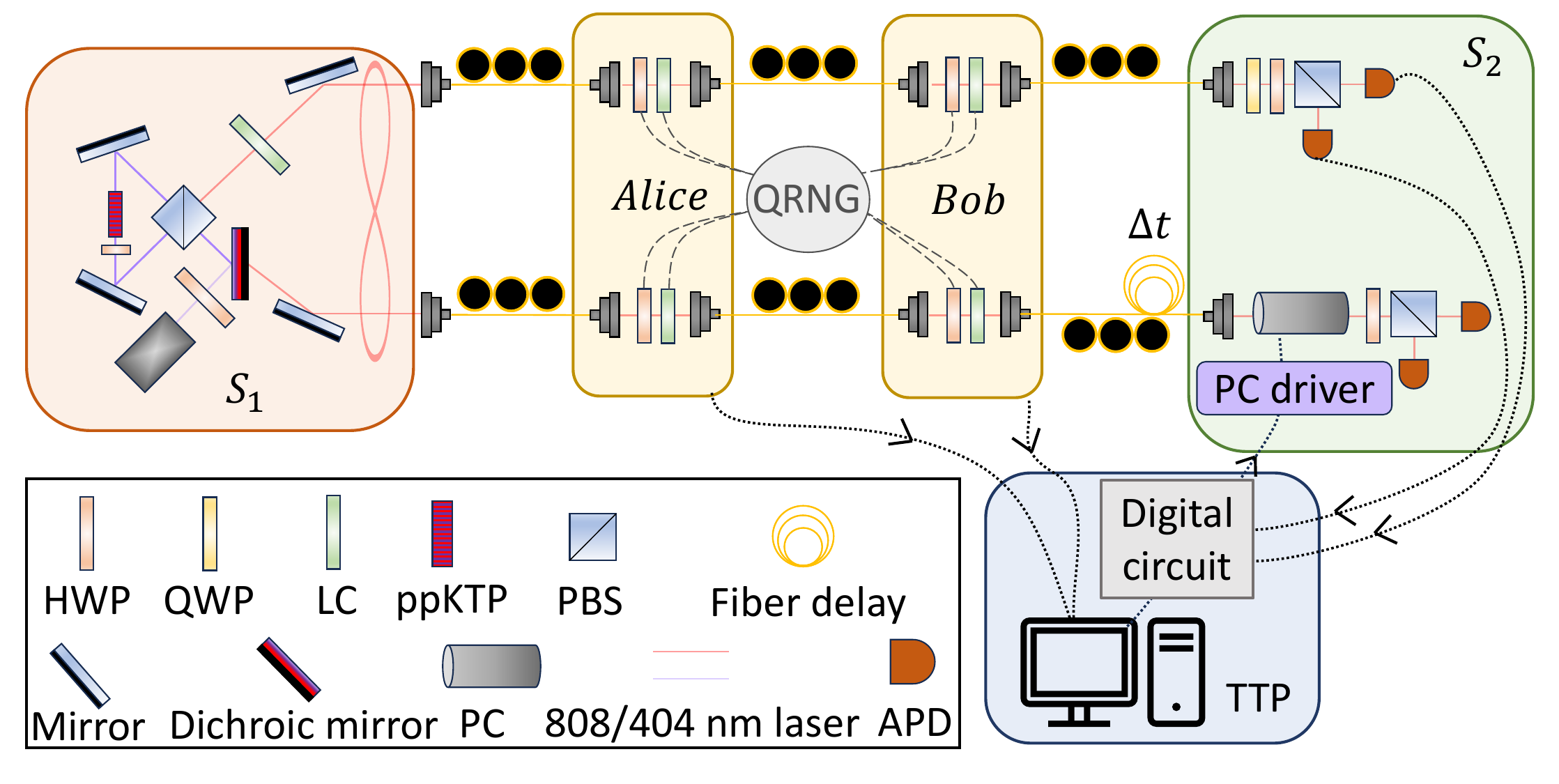}
    \caption{\textbf{Experimental setup.} A Sagnac-based source generates entangled photon pairs in the state in Eq.~\eqref{eq:generated_state}. The pairs are sent via a fiber link to the two clients who perform their transformation through HWPs and LCs. The entangled pair is then sent to server $S_2$, which measures the two qubits. The second qubit measurement is chosen according to the first measurement outcome via an active feed-forward system acting on an electro-optical modulator, i.e. a Pockels cell.} 
    \label{fig:setup}
\end{figure}

\textbf{\textit{Experimental setup.}}
The experimental setup employed is depicted in Fig.~\ref{fig:setup}.
Server S$_1$ is a Sagnac-based source of polarization-entangled photon pairs \cite{fedrizzi2007wavelength}. The qubit state $\ket{0}$ is encoded in the photons' horizontal polarization ($\ket{H}$) and $\ket{1}$ in the vertical one ($\ket{V}$).
The TTP is made up of a computer equipped with an input/output digital module and a fast digital circuit.  Further information about the feed-forward system can be found
in the Supplementary Information. The clients' secret parameters are stored in the TTP computer and used to compute the first measurement basis, $\delta_1$, and the two possible values of the second measurement basis, $\delta_2$ for computation rounds, i.e. $ \delta_2^{\pm} = \theta'_{2} + x_2\pi + r_{2}\pi \pm \phi_2 $.
In test rounds, the second measurement basis is fixed to $ \delta_2^{+}$.
Server S$_2$ is made up of two polarization measurement stations that carry out measurements of the form $M(\delta) = \cos{(\delta)}\sigma_x + \sin{(\delta)}\sigma_y$.
The first measurement station uses a quarter-wave plate (QWP), a half-wave plate (HWP), and a polarizing beam splitter (PBS).
In the second measurement station, the QWP is substituted with a fast electro-optical modulator, i.e. a Pockels cell (PC), which is activated in $\approx 325$~ns with high-voltage pulses, while the second photon is delayed through a $\approx 65$~m single-mode fiber. The avalanche photodiodes (APDs) in the first measurement station are linked to the TTP digital circuit to enable adaptivity of the second measurement basis.
The clients apply their random transformations through half-wave plates rotated at $0^{\circ}$ or $45^{\circ}$, depending on the bit $b_i^j$, and liquid crystals (LC). The clients' random parameters are chosen through an ID Quantique quantum random number generator (QRNG). The QRNG is also used to choose run-by-run between test and computational rounds.

\begin{figure*}
    \centering
    \includegraphics[width=\textwidth]{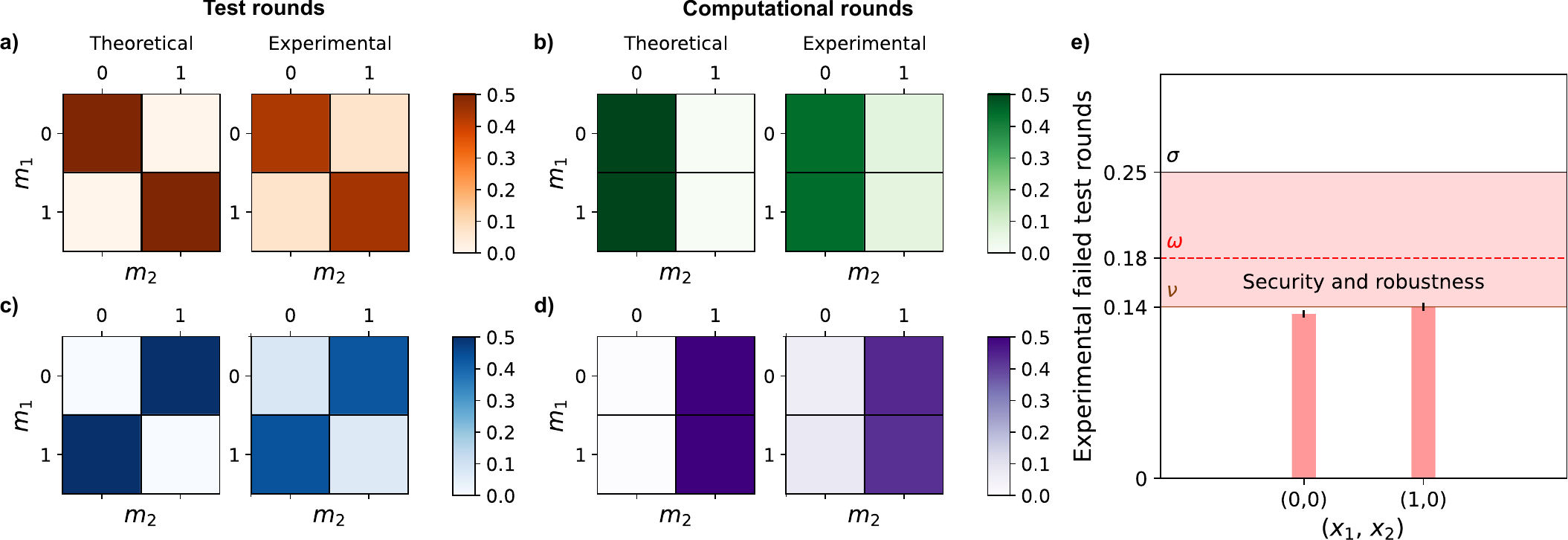}
    \caption{\textbf{Experimental results. a)-b)} Theoretical and experimental outcome probabilities of the test and computation rounds, respectively, for the algorithm $\left(\phi_1, \phi_2,x_1, x_2 \right) =\left(\frac{\pi}{2}, \frac{\pi}{2},0,0 \right) $. \textbf{c)-d)} Theoretical and experimental outcome probabilities of the test and computation rounds, respectively, for the algorithm $\left(\phi_1, \phi_2,x_1, x_2 \right) =\left(\frac{\pi}{2}, \frac{\pi}{2},1,0 \right) $.  \textbf{e)} Robustness and security of the protocol. We find that, for both algorithms, the experimental test error fraction $\epsilon$ is such that $ \epsilon < \omega$. Therefore, the protocol shows robustness and security.}
    \label{fig:results}
\end{figure*}

\textbf{\textit{Results.}}
Our experimental results are shown in Fig.~\ref{fig:results}.
We set the threshold for the tolerated fraction of failed test rounds to $\omega = 0.18$, well in between the lower robustness bound $\nu \approx 0.14$, the error rate expected when considering a noisy model based on the characteristics of the devices we employ, and the upper security bound $\sigma = 0.25$.
This specific choice of $\omega$ is the result of equally minimizing the security error and the robustness error of the protocol, for which concrete expressions can be found in the proofs of Theorems~3 and~4 from~\cite{leichtle2021verifying}, respectively.
The model we define to estimate $\nu$ takes into account several parameters, such as the visibility of the generated entangled state, the errors introduced by the imperfect devices used by the clients, such as waveplates and liquid crystals, and the effective measurement bases on the server's side. Further information about this model are reported in the Supplementary Information.
In Fig.~\ref{fig:results}a-d, we show four different colormaps that summarize the results for the classical computation of two different algorithms. In Fig.~\ref{fig:results}a-b, we show the test and computation rounds, respectively, for the algorithm $\left( \phi_1,\phi_2,x_1,x_2 \right) =\left(\frac{\pi}{2}, \frac{\pi}{2},0,0 \right) $.
Test rounds are used by the clients to draw conclusions on the server behavior.
In particular, since in test rounds the clients measure stabilizers of the two-qubit cluster state employed, the outcomes can only be equal for the two qubits, i.e. one can only have $(m_1,m_2) = (0,0)$ or $(m_1,m_2) = (1,1)$.
We record an experimental error rate equal to $\epsilon_1 = (13.4 \pm 0.3)\%$, as we show in Fig.~\ref{fig:results}a. Since $\epsilon_1 < \omega < \sigma$, the clients can apply the majority vote to get the correct outcome.
The outcomes for the computation rounds related to this algorithm are shown in Fig.~\ref{fig:results}b. The figure shows that in the majority of rounds ($\approx 86.6\%$) the clients obtained the results $(m_1,m_2) = (0,0)$ or $(m_1,m_2) = (1,0)$, which corresponds to the theoretical expectations for this algorithm.
The total number of performed rounds is equal to $27441$, which suppresses the probability that the protocol rejects because of the observed noise below the order of $10^{-26}$.
In all other cases, the clients can conclude that the computation is correct with a soundness error below $10^{-22}$.
These numbers can be obtained from the concrete formulae for the security and robustness errors in the proofs of Theorems~3 and~4 from~\cite{leichtle2021verifying}.
We repeat the experiment for the algorithm $\left(\phi_1, \phi_2,x_1, x_2 \right) =\left(\frac{\pi}{2}, \frac{\pi}{2},1,0 \right) $. In this case, the experimental error rate averaged over all test rounds amounts to $\epsilon_2 = (14.0 \pm 0.3)\%$, which is still compatible with security and robustness.
The outcomes of the test rounds are shown in Fig.~\ref{fig:results}c.
Therefore, also in this case, the clients can apply the majority vote and find that, in the majority of the rounds ($\approx 86\%$), the outcome of the computation is $(m_1, m_2) = (0,1)$ or $(m_1, m_2) = (1,1)$.
The total number of performed rounds for the second algorithm is equal to $24072$, which suppresses the probability that the protocol rejects on a device with the above noise level below the order of $10^{-15}$, with a soundness error below $10^{-19}$.
Also for this algorithm, the results found by the clients correspond to the theoretical expectations, as visible from Fig.~\ref{fig:results}d.
In Fig.~\ref{fig:results}e, we summarize our experimental results by comparing them with the $\omega$ threshold, the maximum expected level of hardware noise $\nu$, and the security threshold $\sigma$.
We conclude that our concrete implementation of the protocol satisfies both the robustness and the security properties to a satisfyingly high degree.\\

\begin{figure}
    \centering
    \includegraphics[width=\columnwidth]{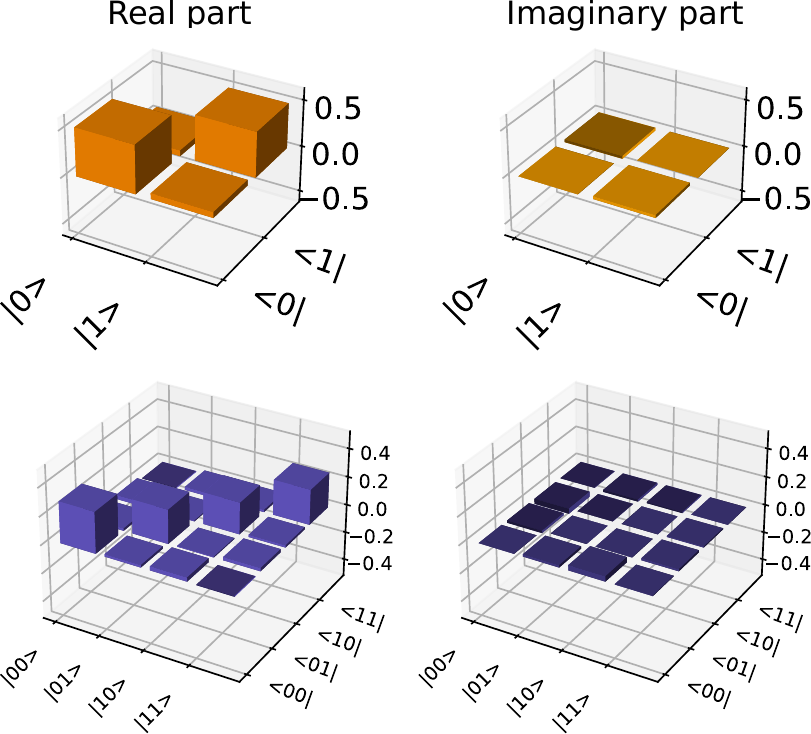}
    \caption{\textbf{Blindness of the state. a)} Density matrix of the second qubit averaged over all possible $\theta_1^{\text{A}}$ and $\theta_1^{\text{B}}$ configurations. \textbf{b)} Density matrix of the two-qubit initial state, averaged over all possible values of $\theta_1^{\text{A}}$ and $\theta_2^{\text{B}}$.}
    \label{fig:blindness}
\end{figure}

We also checked for protocol blindness. In detail, in Fig.~\ref{fig:blindness}a, we show the density matrix obtained as an average over $64$ combinations, namely all possible clients' rotation angles $\theta_1^A$ and $\theta_1^B$, and performing quantum state tomography of the second qubit, while the first qubit is measured in the basis $\delta_1 = 5\pi/4$.
The fidelity with the completely mixed state amounts to $F_{1q} = 0.9952 \pm 0.0003$. In Fig.~\ref{fig:blindness}b, instead, we show the density matrix obtained by averaging over the $64$ density matrices obtained by combining all possible clients' rotation angles $\theta_1^A$ and $\theta_2^B$. The fidelity with the completely mixed state amounts to $F_{2q} = 0.9901 \pm 0.0002$.
To quantify the classical information that the server can retrieve from the received full two-qubit quantum state, we compute the Holevo quantity $\chi$ of the two-qubit state.
Such a quantity is defined as:
\begin{equation}
    \chi = -\Tr{\rho \log_2 \rho} + \sum_i p_i \Tr{\rho_i \log_2 \rho_i}
\end{equation}
where the matrices $$\rho_i = \frac{1}{4}\left( \rho_{\theta_1^A,\theta_2^B} + \rho_{\theta_1^A,\theta_2^B+\pi} + \rho_{\theta_1^A + \pi,\theta_2^B}   + \rho_{\theta_1^A + \pi,\theta_2^B + \pi} \right)$$ are uniformly distributed.
We take the $\rho_i$ density matrices obtained through quantum state tomographies (see also Supplementary Note V), while $\rho$ is obtained by averaging over them.
The experimental value for such a quantity amounts to $\chi_{\text{exp}} = 0.004 \pm 0.001$. This is an upper bound to the amount of classical information that the server can retrieve from the received quantum state.
Our experimental value shows that the server can get almost zero bits of information.

\textbf{\textit{Discussions.} }
In this Letter, we addressed a fundamental open question about the verification of BQC in multi-client scenarios featuring untrusted sources of quantum states.
In particular, we focused on a recently explored protocol and architecture \cite{polacchi2023multi} particularly suitable for multi-client BQC implementations, due to its lightweight structure and the possibility of integrating it into larger networks. 
We tailored such an architecture to the verifiable protocol presented in \cite{kashefi2024verification}, by slightly increasing the quantum capability on the clients' side, who now need to perform not only random $z-$rotations but also bit-flip operations, and by sacrificing some of the protocol runs as a test.
We stress that these extra test rounds introduce only a polynomial overhead in terms of the round complexity of the protocol rather than any qubit counts or change of the platform. Hence the overall protocol remains efficient and scalable similar to the original protocol \cite{polacchi2023multi}, while being equipped with verifiability.
Our results show that the protocol remains blind even after such modifications while being able to provide information about the server's honesty with an arbitrarily high confidence level, depending on the number of runs performed.
Overall, our findings support the first demonstration of a verifiable multi-client BQC.

Equipping clients with tools to verify delegated quantum computations while reducing hardware assumptions has a key role in the roadmap toward a secure quantum internet, a fact that is demonstrated by the huge effort put into the development of device-independent protocols \cite{pironio2016focus,vsupic2020self}.
Dropping assumptions becomes even more difficult for delegated tasks, where the users of quantum devices cannot observe what happens \textit{in situ} on the server side. 
On the other hand, reducing the hardware capacity on the clients' side is also crucial, since access to single qubit sources is not straightforward, especially when considering very sophisticated apparatuses, such as quantum dot sources, requiring cryogenic temperatures.
This context strongly motivates the theoretical investigation and the experimental demonstration of secure and verifiable BQC protocols in scenarios where the clients do not have control over the quantum sources of states as well as the remote quantum computational server node.
Therefore, we believe that our findings represent a significant step towards implementing safe and densely connected quantum networks.

\textbf{\textit{Acknowledgments.}}
All authors acknowledge the support of the European Union’s Horizon 2020 research and innovation program through the FET project PHOQUSING (``PHOtonic Quantum SamplING machine'' – Grant Agreement No. 899544) and by ICSC – Centro Nazionale di Ricerca in High Performance Computing, Big Data and Quantum Computing, funded by European Union – NextGenerationEU. B.P. acknowledges support from the Sapienza research grant Avvio alla Ricerca 2022 (No. AR1221816B79122A). D.L. and E.K. acknowledge the support of the ANR research grant ANR-21-CE47-0014 (SecNISQ) and the EPSRC grant EP/X026167/1.
D.L. acknowledges support by the Engineering and Physical Sciences Research Council [grant reference EP/T001062/1], and the UK National Quantum Computer Centre [NQCC200921], which is a UKRI Centre and part of the UK National Quantum Technologies Programme (NQTP).

\end{document}


\title{Supplementary Information: Experimental verifiable multi-client blind quantum computing on a Qline architecture}

\author{Beatrice Polacchi}
\affiliation{Dipartimento di Fisica - Sapienza Universit\`{a} di Roma, P.le Aldo Moro 5, I-00185 Roma, Italy}

\author{Dominik Leichtle}
\affiliation{Laboratoire d’Informatique de Paris 6, CNRS, Sorbonne Université, 75005 Paris, France}

\author{Gonzalo Carvacho}
\affiliation{Dipartimento di Fisica - Sapienza Universit\`{a} di Roma, P.le Aldo Moro 5, I-00185 Roma, Italy}

\author{Giorgio Milani}
\affiliation{Dipartimento di Fisica - Sapienza Universit\`{a} di Roma, P.le Aldo Moro 5, I-00185 Roma, Italy}

\author{Nicol\`o Spagnolo}
\affiliation{Dipartimento di Fisica - Sapienza Universit\`{a} di Roma, P.le Aldo Moro 5, I-00185 Roma, Italy}

\author{Marc Kaplan}
\affiliation{VeriQloud, 13 rue Victor Hugo, 92 120 Montrouge, France}

\author{Elham Kashefi}
\email{elham.kashefi@lip6.fr}
\affiliation{Laboratoire d’Informatique de Paris 6, CNRS, Sorbonne Université, 75005 Paris, France}
\affiliation{School of Informatics, University of Edinburgh, 10 Crichton Street, EH8 9AB Edinburgh, United Kingdom}

\author{Fabio Sciarrino}
\email{fabio.sciarrino@uniroma1.it}
\affiliation{Dipartimento di Fisica - Sapienza Universit\`{a} di Roma, P.le Aldo Moro 5, I-00185 Roma, Italy}

\maketitle

\section{Noise model}
\label{sec:noise_model}

In the case of an honest source of states, the state generated by the Sagnac source is the following:

\begin{equation}
    \ket{\psi^+} = \frac{1}{2} \left( \ket{HH}  + \ket{VV} \right) 
\end{equation}

We modeled the experimental state with white and colored noise, according to the following formula:

\begin{equation}
    \rho_{noisy} = v \ket{\psi^+}\bra{\psi^+} + (1-v) \left[ \frac{\lambda}{2}~( \ket{\psi^+}\bra{\psi^+} + \ket{\psi^-}\bra{\psi^-} ) + \frac{1-\lambda}{4} \mathbb{I}_4 \right ]
\end{equation}
where $v$ is the visibility of the state, $l$ is the fraction of colored noise, and $\ket{\psi^-} = \frac{1}{2} \left( \ket{HH}  - \ket{VV} \right)$.

The parameters $v$ and $\lambda$ of the generated state are compatible with values $v \approx 0.935$ and $\lambda \approx 0.493$. The error on the liquid crystals amounts, on average, to  $\approx \pi/16$ in absolute value. 
The noise we consider on the half-wave plates of the server's measurement stations, instead, ranges in the interval $[-1^{\circ}, 1^{\circ}]$, on average.
We also consider an imperfect phase shift inserted by the Pockels cell, having an error amounting to $\approx \pi/16$.

\section{Feed-forward system}
\label{sec:adaptivity}

We describe here our feed-forward system for measurement adaptivity.

The measurement operator applied in the second measurement station is the following:
\begin{equation}
    O(\delta_2) = 
    X^f~U^{\dagger}_{PC}(\phi)~U^{\dagger}_{HWP}(\pi/8)~O_{PBS}~U_{HWP}(\pi/8)~U_{PC}(\phi)~X^f
\label{eq:O_delta2}
\end{equation}

The values of $\delta_2$, the phase shift $\phi$ inserted by the Pockels cell (PC), and the bit-flip $f$, which is applied to the second outcome in post-processing, are summarized in Supplementary Table~\ref{tab:phase-shifts}.

\begin{table}[h!]
\centering
\begin{tabular}{cccc}
$\boldsymbol{\delta_2}$ & \textbf{Ideal phase-shift} & \textbf{PC phase-shift} & \textbf{f} \\
0                   & 0                          & 0                       & 0                    \\
$\pi/4$             & $7\pi/4$                   & $3\pi/4$                & 1                   \\
$\pi/2$             & $3\pi/2$                   & $\pi/2$                 & 1                   \\
$3\pi/4$            & $5\pi/4$                   & $\pi/4$                 & 1                   \\
$\pi$               & $\pi$                      & 0                       & 1                   \\
$5\pi/4$            & $3\pi/4$                   & $3\pi/4$                & 0                    \\
$3\pi/2$            & $\pi/2$                    & $\pi/2$                 & 0                    \\
$7\pi/4$            & $\pi/4$                    & $\pi/4$                 & 0                   
\end{tabular}
\caption{}
\label{tab:phase-shifts}
\end{table}

At the core of the feed-forward system, there is the digital circuit depicted in Supplementary Figure ~\ref{fig:fast data elaboration}. It gets a ten-bit input and produces a five-bit output according to a logic function $h(A,B,r_1,m_1^+,m_1^-,c)=V,m_1^{+,\mathrm{true}},m_1^{-,\mathrm{true}}$.\\

\textbf{INPUT:}
\begin{itemize}
    \item the three-bit input $A=a_2a_1a_0 $ encodes the angle $\theta'_2+x_2\pi +r_2 \pi$;
    \item the three-bit input $B=b_2b_1b_0$ encodes the angle $\phi_2=\pi/2$;
    \item the input bit $r_1 = r_1^{\text{A}}\oplus r_1^{\text{B}}$ is used to correct the value of $m_1$;
    \item the two-bit input $\{m_1^+,~m_1^- \}\in\{10,01\}$ is the physical signal of the two detectors in the first measurement station and encodes, respectively, the logical outcomes $m_1=0,1$;
    \item the bit $c$ decides whether to implement a computational or test round.
\end{itemize}

\textbf{OUTPUT:}
\begin{itemize}
    \item the three-bit output $V=v_2v_1v_0$ encodes four different voltage amplification factors for the implementation of the phase-shifts reported in Supplementary Table~\ref{tab:phase-shifts};
    \item the bits $m_1^{+\mathrm{true}},~m_1^{-\mathrm{true}}\in\{10,01\}$ encode the corrected first logical outcome $m_1^{\mathrm{true}}=0,1$ and are sent to a coincidence box for data analysis, together with the second measurement outcome. Finally, the second outcome is also corrected in post-processing as $m_2^{\mathrm{true}} = m_2 \oplus r_2 \oplus f$.
\end{itemize}

The function $h$ is such that the three-bit output $V=v_2v_1v_0$ encodes the value of $\delta_2$ depending on whether a test ($c=1$) or computational round ($c=0$) is being performed. In detail, the output $V=v_2v_1v_0$ sets the measurement basis $(1-c) (A + (-1)^{m_1^{\mathrm{true}} } B ) + c (A+B)$. All values of $A,~B,~f,~V$ are summarized in Supplementary Table~\ref{tab:FDE-encodings}.

\begin{table}[h!]
\centering
\begin{tabular}{llrl}
$a_2a_1a_0\leftrightarrow A[\text{rad}]\quad$ & $b_2b_1b_0\leftrightarrow B[\text{rad}]\quad$ & $\delta_2\leftrightarrow fv_2v_1v_0$ & $\leftrightarrow$\textbf{PC voltage} \\
$\quad000\qquad0$                             & $\quad000\qquad0$                             & $0\qquad0000\quad$                   & $\qquad\text{V}_0$ \\
$\quad001\qquad\pi/4$                         & $\quad001\qquad\pi/4$                         & $\pi/4\qquad1100\quad$               & $\qquad\text{V}_{3\pi/4}$ \\
$\quad010\qquad\pi/2$                         & $\quad010\qquad\pi/2$                         & $\pi/2\qquad1010\quad$               & $\qquad\text{V}_{\pi/2}$ \\
$\quad011\qquad3\pi/4$                        & $\quad011\qquad3\pi/4$                        & $3\pi/4\qquad1001\quad$              & $\qquad\text{V}_{\pi/4}$ \\
$\quad100\qquad\pi$                           & $\quad100\qquad\pi$                           & $\pi\qquad1000\quad$                 & $\qquad\text{V}_0$ \\
$\quad101\qquad5\pi/4$                        & $\quad101\qquad5\pi/4$                        & $5\pi/4\qquad0100\quad$              & $\qquad\text{V}_{3\pi/4}$ \\
$\quad110\qquad3\pi/2$                        & $\quad110\qquad3\pi/2$                        & $3\pi/2\qquad0010\quad$              & $\qquad\text{V}_{\pi/2}$ \\
$\quad111\qquad7\pi/4$                        & $\quad111\qquad7\pi/4$                        & $7\pi/4\qquad0001\quad$              & $\qquad\text{V}_{\pi/4}$
\end{tabular}
\caption{}
\label{tab:FDE-encodings}
\end{table}

\begin{figure}[ht!]
    \centering
    \includegraphics[width=0.6\textwidth]{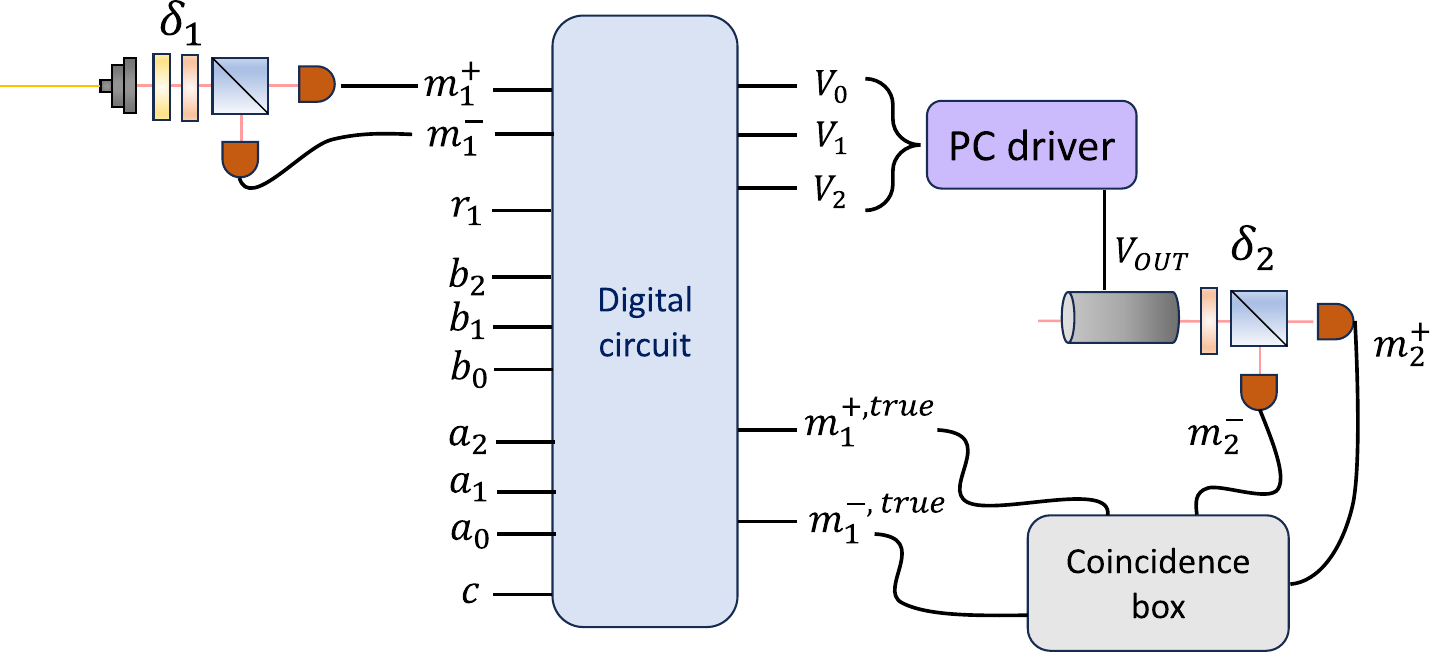}
    \caption{\textbf{Feed-forward system.} A digital circuit gets as input: (i) three bits corresponding to the angle $A = \theta'_2+x_2\pi +r_2 \pi $, (ii) three bits corresponding to the angle $B = \phi_2 = \pi/2$, (iii) one bit for $r_1$, (iv) the two detector signals corresponding to $m_1^+$ and $m_1^-$, and (v) the bit $c$ indicating whether a test or computational round is performed. The function corrects the outcome of the first measurement and evaluates $\delta_2$ accordingly. The digital circuit gives the following output: (i) three bits, $V_0,~V_1,~V_2$, corresponding to three different voltage amplification factors applied to the PC by an analog PC driver, (ii) a copy of the two corrected signals $m_{1}^{\pm,\mathrm{true}}$. All outcomes are sent to a coincidence box for data analysis.}
    \label{fig:fast data elaboration}
\end{figure}

\newpage
\section{Temporal scheme of the experiment}
\label{sec:time_scheme}

Each round of the protocol is divided into three stages:

\begin{itemize}
    \item At time $t_0$, the clients set their half-wave plates (HWPs) and liquid crystals (LCs) according to the random secret transformation they want to apply to the qubits. They deliver their secret parameters to the TTP. Such parameters are stored in the TTP computer and used to compute the first measurement angle, $\delta_1$, and the two possible values for $\delta_2^{\pm}$ in computation rounds, or $\delta_2^{+}$ in test rounds. The TTP communicates the angle $\delta_1$ to server S$_2$ that sets the first measurement station accordingly;

    \item At time $t_1$, the photon pairs are generated and sent through Alice's and Bob's HWPs and LCs. Then, S$_2$ measures the first photon, while the second one is delayed through a $65~$m single-mode fiber loop. The outcome of the first measurement is sent to the TTP digital circuit for correction and final computation of $\delta_2$;

    \item At time $t_2$, the TTP communicates the voltage to be applied to the PC to measure the second qubit in the $\delta_2$ basis, and the second photon is also measured. The second outcome is given back to the TTP for correction.
\end{itemize}

\begin{figure}[htb!]
    \centering
    \includegraphics[width=\textwidth]{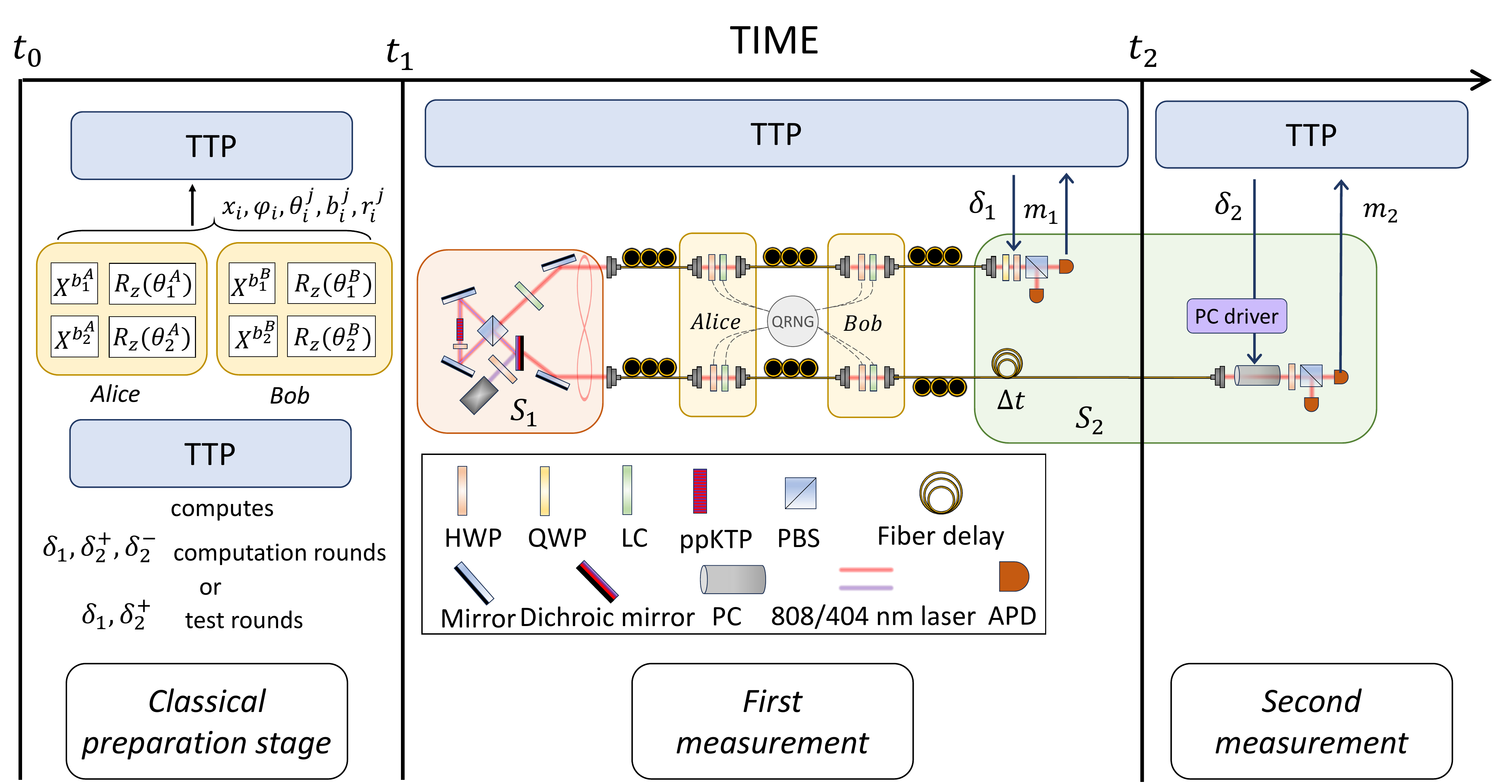}
    \caption{\textbf{Time scheme of the experiment.} The temporal scheme of the experiment is divided into three phases: at $t_0$ the classical preparation stage starts, during which the clients set up their waveplates and liquid crystals and communicate their secret parameters to the TTP. At $t_1$, the quantum part of the protocol starts and the first measurement takes place. At $t_2$, finally, the second measurement is performed.}
    \label{fig:time_scheme}
\end{figure}

\newpage

\section{Quantum state tomographies of the second qubit}
\label{sec:blindness_second}

In this section, we show eight of the quantum state tomographies used to compute the blindness of the second qubit, with the corresponding fidelity with the ideal state.
For all density matrices, we fixed the measurement angle $\delta_1 = 5\pi/4$.

\begin{figure}[ht!]
    \centering
    \includegraphics[width=0.9\textwidth]{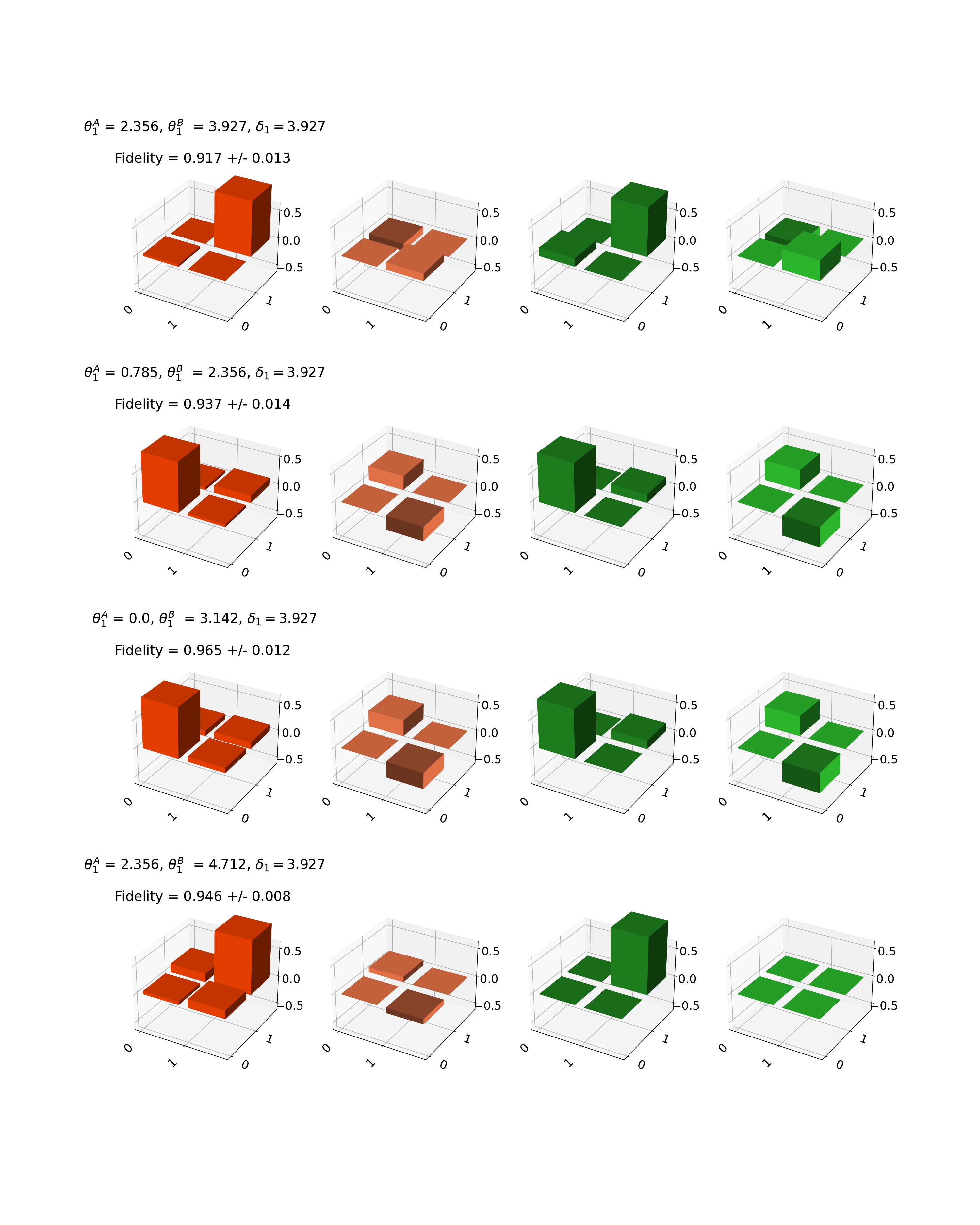}
\end{figure}

\begin{figure}[ht!]
    \centering
    \includegraphics[width=0.9\textwidth]{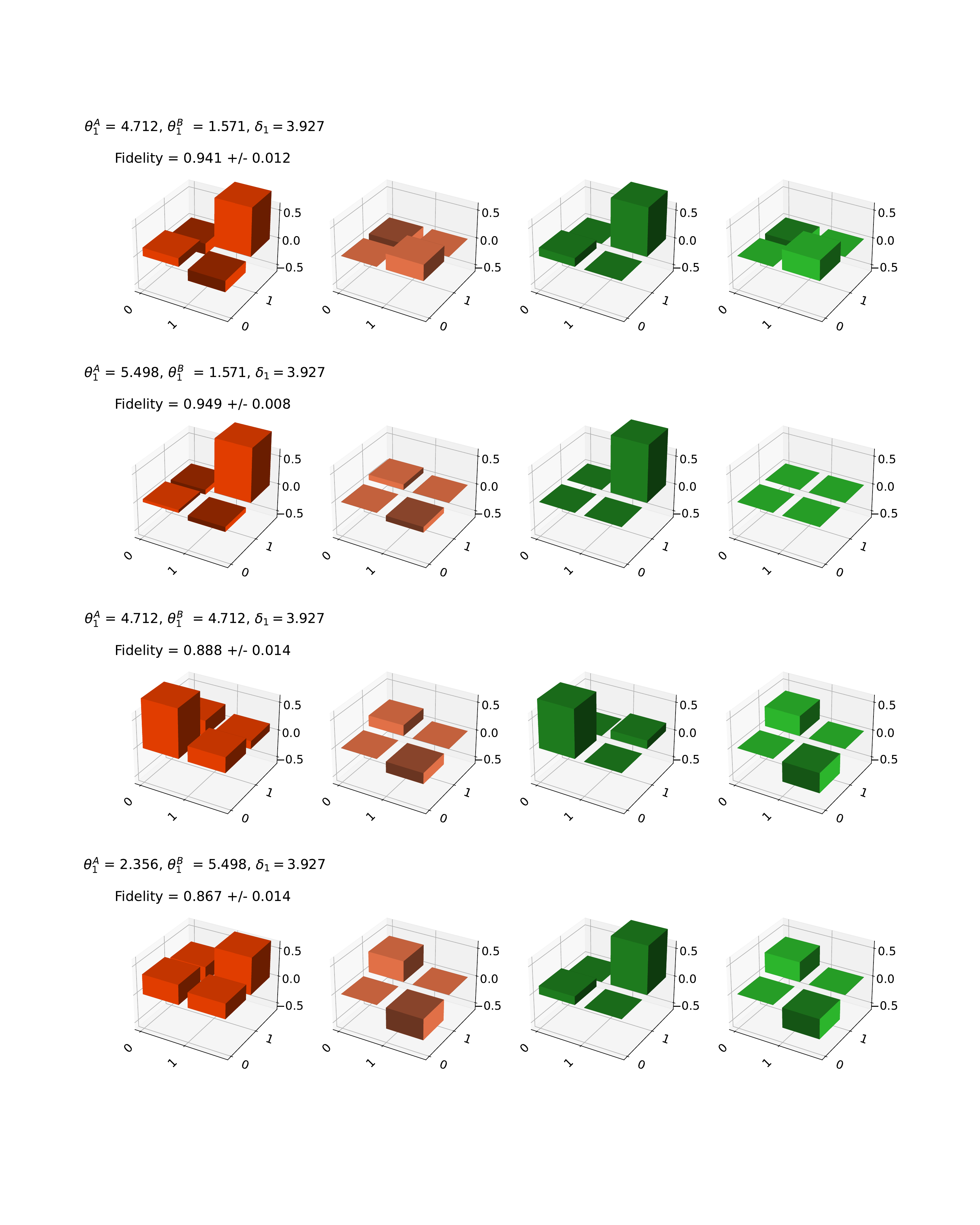}
    \caption{\textbf{Density matrices of the second qubit.} We show eight randomly chosen experimental and theoretical quantum state tomographies of the second qubit when the first measurement angle was set to $\delta_1 = 5\pi/4$. In orange, we report, respectively, the real and imaginary parts of the experimental density matrix, while in green the real and imaginary parts of the ideal density matrix. We also report the corresponding fidelities.}
\end{figure}

\newpage

\section{Quantum state tomographies of the full initial state}
\label{sec:blindness_both}

In this section, we show eight of the quantum state tomographies used to compute the blindness of the full initial two-qubit state, with the corresponding fidelity with the ideal states.

\begin{figure}[ht!]
    \centering
    \includegraphics[width=0.95\textwidth]{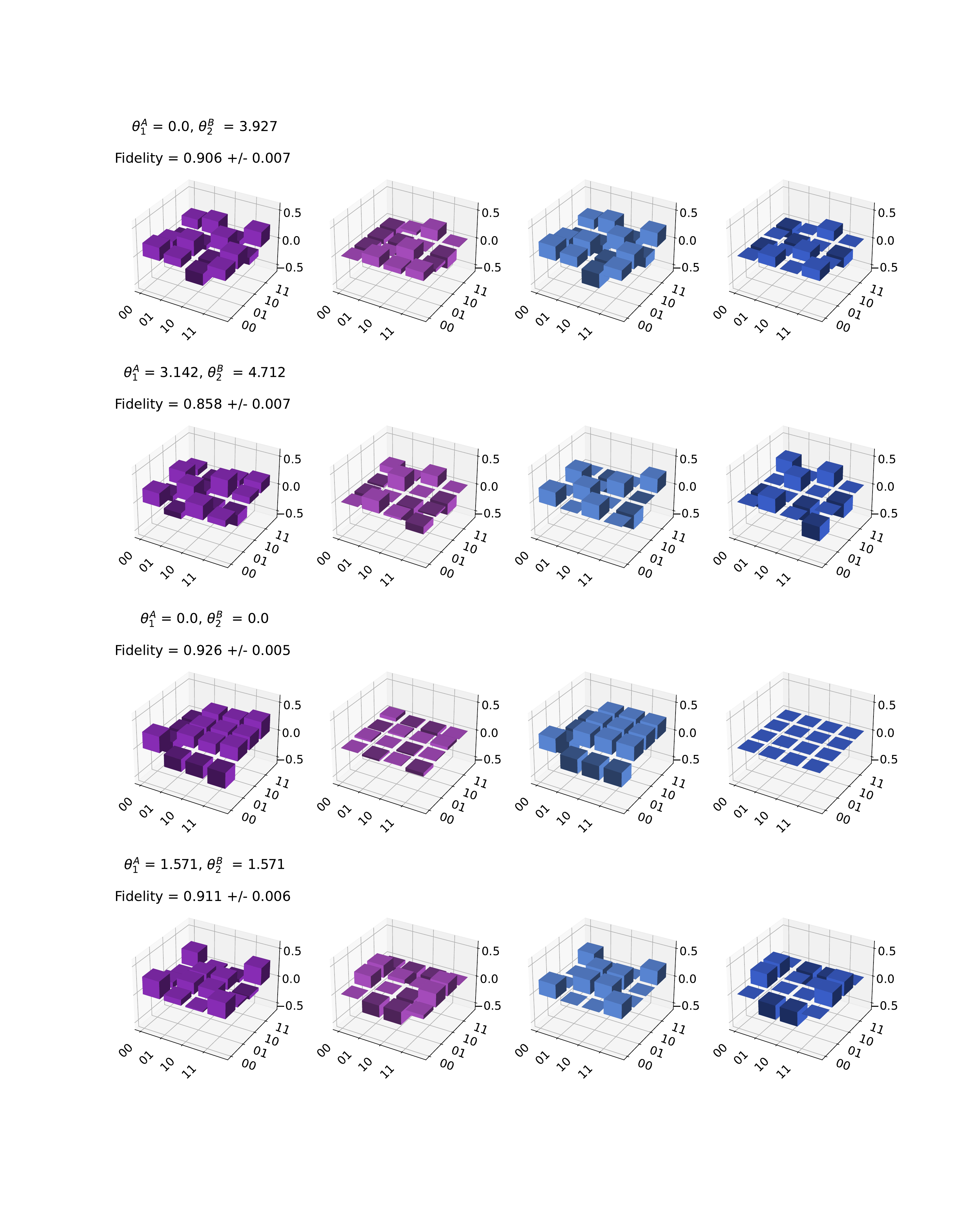}
\end{figure}

\begin{figure}[ht!]
    \centering
    \includegraphics[width=\textwidth]{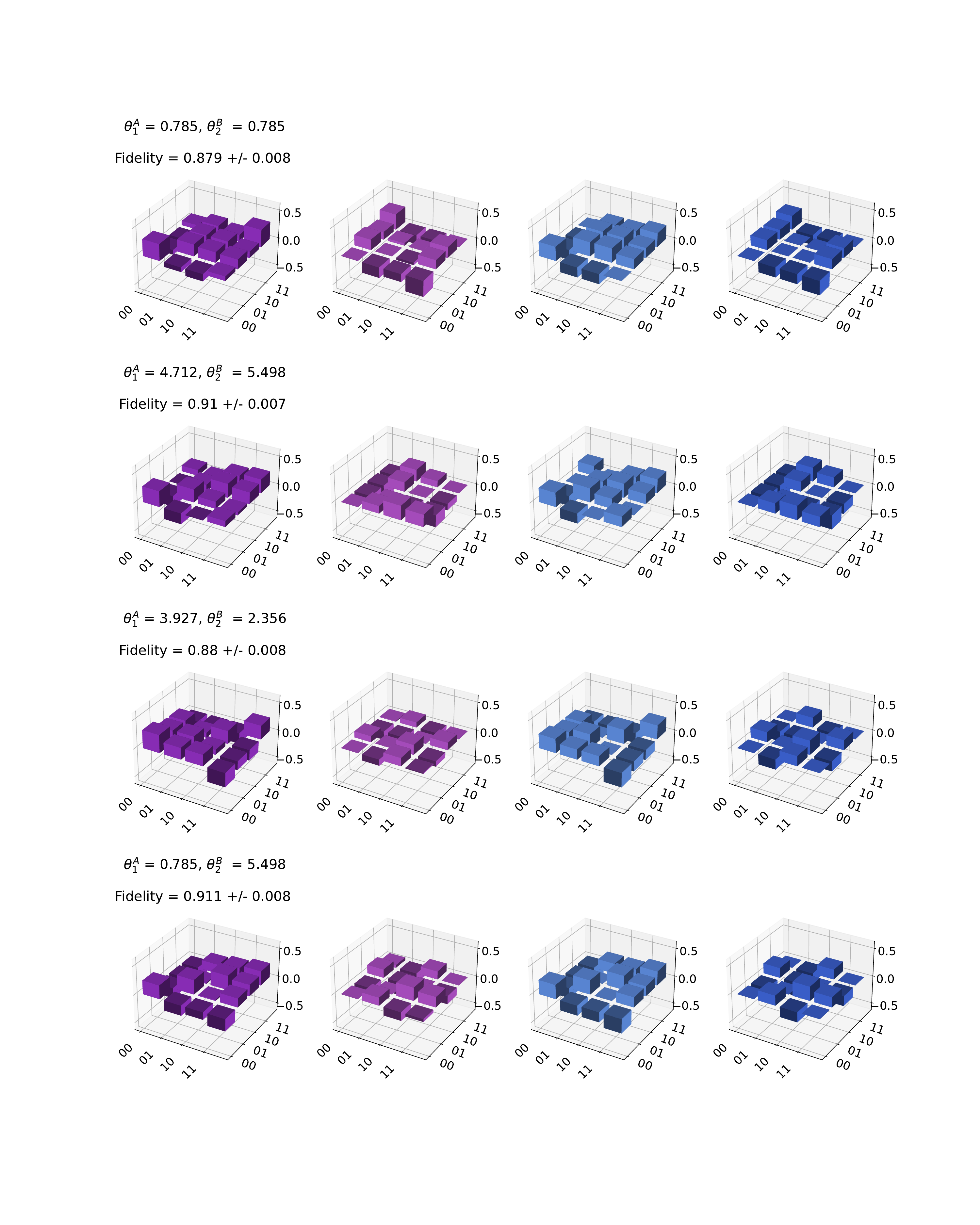}
    \caption{\textbf{Density matrices of the full initial state.} We show eight randomly chosen experimental and theoretical quantum state tomographies of the full two-qubit state. In purple, we report, respectively, the real and imaginary parts of the experimental density matrix, while in blue the real and imaginary parts of the ideal density matrix. We also report the corresponding fidelities.}
\end{figure}